%% file: www.tex
\documentclass[sigconf]{acmart}
\usepackage{multirow}
\usepackage{graphicx}
\usepackage{float}
\usepackage{colortbl}
\usepackage{subfig}
\usepackage{enumitem}
\AtBeginDocument{%
  }

\copyrightyear{2025}
\acmYear{2025}
\setcopyright{acmlicensed}\acmConference[WWW Companion '25]{Companion Proceedings of the ACM Web Conference 2025}{April 28-May 2, 2025}{Sydney, NSW, Australia}
\acmBooktitle{Companion Proceedings of the ACM Web Conference 2025 (WWW Companion '25), April 28-May 2, 2025, Sydney, NSW, Australia}
\acmDOI{10.1145/3701716.3715596}
\acmISBN{979-8-4007-1331-6/2025/04}

\begin{document}

\title{Leveraging Memory Retrieval to Enhance LLM-based Generative Recommendation}


\author{Chengbing Wang} 
\email{wwq197297@mail.ustc.edu.cn}
\affiliation{%
  \institution{University of Science and Technology of China}
  \city{Hefei}
  \country{China}
}

\author{Yang Zhang} 
\email{zyang1580@gmail.com}
\affiliation{%
  \institution{National University of Singapore}
  \city{Singapore}
  \country{Singapore}
}
\authornote{Corresponding author}

\author{Fengbin Zhu} 
\email{zhfengbin@gmail.com}
\affiliation{%
  \institution{National University of Singapore}
  \city{Singapore}
  \country{Singapore}
}
\authornotemark[1]

\author{Jizhi Zhang} 
\email{cdzhangjizhi@mail.ustc.edu.cn}
\affiliation{%
  \institution{University of Science and Technology of China}
  \city{Hefei}
  \country{China}
}

\author{Tianhao Shi} 
\email{sth@mail.ustc.edu.cn}
\affiliation{%
  \institution{University of Science and Technology of China}
  \city{Hefei}
  \country{China}
}

\author{Fuli Feng} 
\email{fulifeng93@gmail.com}
\affiliation{%
  \institution{University of Science and Technology of China}
  \city{Hefei}
  \country{China}
}





\renewcommand{\shortauthors}{Chengbing Wang et al.}

\begin{abstract}
\input{main/0_abstract}

\end{abstract}




\maketitle

\input{main/1_intro}
\input{main/4_methods}
\input{main/5_experiments}

\input{main/6_conclusion}
\begin{acks}
The numerical calculations in this paper have been done on the supercomputing system in the Supercomputing Center of University of Science and Technology of China. This research was also supported by the advanced computing resources provided by the Supercomputing Center of the USTC.
\end{acks}

\bibliographystyle{ACM-Reference-Format}
\balance
\bibliography{www}



\end{document}

%% file: main/0_abstract.tex
Leveraging Large Language Models (LLMs) to harness user-item interaction histories for item generation has emerged as a promising paradigm in generative recommendation.
However, the limited context window of LLMs often restricts them to focusing on recent user interactions only, leading to the neglect of long-term interests involved in the longer histories.
To address this challenge, we propose a novel \textit{Automatic Memory-Retrieval} framework (\textbf{\textit{AutoMR}}), which is capable of storing long-term interests in the memory and extracting relevant information from it for next-item generation within LLMs.
Extensive experimental results on two real-world datasets demonstrate the effectiveness of our proposed AutoMR framework in utilizing long-term interests for generative recommendation.

%% file: main/1_intro.tex
\section{Introduction}

Leveraging Large Language Models (LLMs) for generative recommendations is emerging as a promising paradigm in recommendation systems~\cite{LLMRecSurvey,bigrec,d3,lc-rec,GNR}. Unlike traditional methods, this approach fine-tunes LLMs to directly generate items as recommendations based on users' historical interactions, eliminating the need for predefined candidate items~\cite{d3,gpt4rec,bigrec}. Longer histories typically provide a more comprehensive representation of user interests~\cite{SIM}. However, due to computational limitations and restricted context windows, current LLM-based generative recommendation methods often fail to effectively utilize long histories~\cite{bigrec}, potentially leading to suboptimal performance.

To address this issue, the "memorize-then-retrieve" framework offers a promising solution by storing users' extensive historical interactions in memory and retrieving relevant portions when needed for LLM use. However, the retrieval process presents a significant challenge. Since candidate items are no longer provided, retrieval must rely on other local input information, such as recent historical interactions, to extract relevant historical data for the LLM. In this context, simple retrieval methods (e.g., semantic retrieval) often fail to provide meaningful results. For example, when local input information is insufficient for generating accurate recommendations, semantic retrieval may retrieve results that resemble the local data, rather than introducing new insights. Additionally, it is challenging to heuristically identify which aspects of long-term interests are relevant to a specific scenario, complicating the manual design of effective retrieval strategies.

While designing an effective retrieval mechanism manually is challenging, annotating the usefulness of each memory element is relatively straightforward~\cite{Hamed_Zamani}. 
These annotations naturally capture patterns indicating which historical information is valuable. Leveraging this data, we can automatically optimize the retrieval process to learn an effective mechanism. 
Building on this idea, we propose a new \textit{Automatic Memory-Retrieval} (AutoMR) framework that learns to retrieve effectively. AutoMR stores users' long histories, encoded by LLMs, in external memory and annotates each memory sample based on how its inclusion in the LLM prompt reduces the predicted perplexity (PPL) of the ground-truth item. It then leverages these annotations to train the retriever to focus on the most useful information when retrieving.

The main contributions are summarized as follows:
\begin{itemize}[leftmargin=*]
    \item We highlight the importance of modeling long-term history for generative LLM-based recommendation and emphasize the potential of the memory-retrieval framework as a solution.
    
    
    \item  We propose AutoMR, a novel memory-retrieval framework that enables automatic learning of how to extract useful information from memory to support next-item generation within LLMs.
    
    \item We conduct extensive experiments on two widely used datasets, showing that AutoMR effectively improves the ability of LLMs to leverage long-term interests for generative recommendation.
\end{itemize}

%% file: main/4_methods.tex
\section{Preliminaries}
\label{Preliminaries}
Let $\mathcal{D}$ denote the user-item interaction data, where each sample in $\mathcal{D}$ is represented as $(u, y, t)$, indicating that user $u$ interacted with item $y$ at time point $t$. Our goal is to leverage this data to train a recommender model.

\textit{LLM-based Generative Recommendation.} To develop a generative recommender model using LLMs, for each sample $(u, y, t)$, we typically aggregate the most recent $N$ items interacted with by the user $u$ to represent his/her preference, denoted as $h_t$. The data is then transformed into a textual instruction, represented as $x_{u, h_t}$, prompting the LLM to generate a recommendation. For example, the instruction might be: \textit{``A user has interacted with [Item title1, ..., Item title n]. Please predict the next item the user is likely to engage with:"}  
When training, the item $y$ is used as the ground-truth next item, and the LLM is tuned by optimizing the causal language modeling loss. Formally,  
\begin{equation}\label{eq1}\small 
\max\sum_{(u,y,t)\in\mathcal{D}}\sum_{i=1}^{|y|}\log(LLM(y_i|x_{u,h_t},y_{<i})),  
\end{equation}  
where $|y|$ is the total number of tokens in $y$, $y_i$ denotes the $i$-th token, and $LLM(y_i|x_{u,h_t}, y_{<i})$ is the probability predicted by the LLM based on the instruction $x_{u, h_t}$ and all preceding tokens in $y$, \textit{i.e.}, $y_{<i}$. At inference, the trained LLM generates a recommended item directly based on the provided instruction, with an optional grounding process to map the generated item to actual items~\cite{bigrec}.

Existing work often limits the length of $h_{t}$ due to the computational cost and the constrained context window of LLMs, which restricts the utilization of long-term user preference information. This work focuses on addressing this limitation by leveraging the memory-retrieval mechanism.

\section{AutoMR}
\begin{figure}[tbp]
  \centering
  \includegraphics[width=\linewidth]{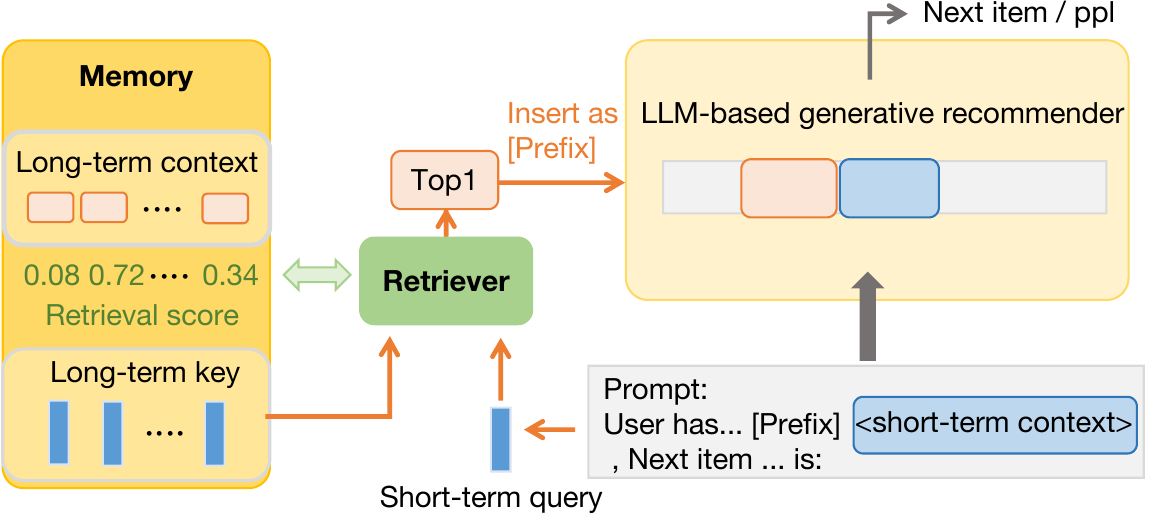}
  \caption{An overview of the proposed AutoMR, which includes three key components: Memory, Retriever, and LLM-based generative recommender (e.g., BIGRec).
  }
  \label{framework}
\end{figure}

In this section, we present our \textit{Automatic Memory-Retrieval} (AutoMR) framework for generative recommendation.

\subsection{Framework Overview} Figure~\ref{framework} shows the overall framework of our AutoMR, the core of which is to externally connect a memory and retriever module to the LLM-based generative recommender. The memory stores the encoded long-term interactions, and the retriever is used to extract useful historical interactions from the memory. 

\noindent \textbf{Memory}. For a user at time point $t$, the memory would store the user all interactions till $t$, except for the most recent $K$ interactions $h_{t}$. To facilitate retriever learning and minimize repeated computation costs, for each interaction needed to store, we directly store its hidden representations obtained in the $L$-th layer of the generative LLM recommender. Specifically, the memory is defined as:  
\begin{equation}\small
    M_{u,t} = [z_1, \dots, z_m, \dots, z_{t - |h_t|}],
\end{equation}
where $z_m=LLM^{(:L)}(x_{u,h_{m}})$ corresponds to the $m$-th interaction's hidden representation obtained at $L$-th layer of the generative recommender, similarly for others. 

\noindent \textbf{Retriever.} Given a user $u$ at time $t$ and his/her most recent interaction $h_t$, the retriever would extract the most useful information from the memory to help the next item prediction. In the process, the query is defined as the local user information, encoding to the hidden representation at the $L$-th layer output, denoted as $z_{t} = LLM^{(:L)}(x_{u,h_t})$.  
The retriever is implemented using an MLP layer, which takes $z_{t}$ and the memory $M_{u,t}$ as inputs and would output a relevance score to each element in $M_{u,t}$. Formally,  
\begin{equation}\label{eq:scores}  \small
    \begin{split}  
        S &= f_{\theta}(z_{t}; M_{u,t}), \\  
        s_m &= softmax\big(MLP(z_t)^{\top}M_{u,t}\big)_{m},  
    \end{split}  
\end{equation}  
where $\theta$ represents the learnable parameters. The retriever then selects the element $z_{m^\prime}$ from $M_{u,t}$ with the highest relevance score.  

\noindent\textbf{Recommendation.} For the final recommendation, we generally adhere to the inference process of existing LLM-based generative recommenders. However, we would combine the retrieved result with the local information to produce the final recommendation. Formally, the recommendation is generated as follows:  
\begin{equation}\label{eq:rec} \small 
    LLM^{(L:)}(combine(z_t, z_{m^{'}})),  
\end{equation}  
where $LLM^{(L:)}$ represents the LLM excluding the bottom $L$ layers.

\subsection{Retriver Training}
Since our memory and retriever module is external to the LLM-based generative recommender, we can directly utilize a well-tuned generative recommender and focus solely on tuning the memory and retriever module. Since the memory itself does not involve learnable parameters, we only need to train the retriever. To achieve this, we first manually annotate memory data based on their usefulness for recommendations, and use these annotations to train the retriever, enabling it to automatically learn to extract relevant information from the memory. The process consists of two main steps: 1) Data annotation, and 2) Training.

\noindent \textbf{Data Annotation} Given a sample $(u, y, t)$ and the user's most recent interactions $h_t$, we assign a relevance score label for each element $z_m$ in the corresponding memory $M_{u,t}$ by combining it with local information to assess whether incorporating it enhances the LLM prediction for $y$ --- whether it reduces the perplexity of predicting $y$. The label is determined based on the change in perplexity when the element is included versus when it is excluded. Formally, the label $s_m^{\prime}$ for $z_m$ is computed as follows:
\begin{equation}\small
\begin{split}
    s_m^{\prime} =  & \text{softmax}([\delta_1, \dots, \delta_m, \dots])_{m} \\
    \delta_m = & \prod_{i} LLM^{(L:)}(y_i | z_t, z_m; y_{<i}) - \prod_{i} LLM^{(L:)}(y_i | z_t; y_{<i}),
\end{split}
\end{equation}
where $LLM^{(L:)}(y_i | z_t, z_m; y_{<i})$ denotes the predicted probability for the $i$-th token in $y$ when combining $z_m$ with the encoded local user information $z_t$ at the $L$-th layer. Similarly, $LLM^{(L:)}(y_i | z_t; y_{<i})$ represents the predicted probability using only the local information. The labels for all memory elements are represented as $S = [s_1^{\prime}, \dots, s_m^{\prime}, \dots]$.

\noindent\textbf{Training}. After obtaining the annotated data, where each sample is represented as $(u, t, h_t, M_{u,t}, S^{\prime})$, we use it to train the retriever to produce scores ($S$, as defined in Equation~\eqref{eq:scores}) that align with the annotated labels. Formally, The retriever is trained by optimizing the following objective:  
\begin{equation}\small
    \min_{\theta} \sum_{(u, t, h_t, M_{u,t}, S^{\prime})} KL\text{Loss}(S^{\prime}, S),
\end{equation}  
where $KL\text{Loss}(\cdot)$ represents the KL divergence loss~\cite{KL}.

\section{Related Work}
Inspired by the generative power of LLMs, great efforts~\cite{LLMRecSurvey, li_survey, bigrec,d3,collm,Text-like} have been made to utilize them as recommenders.
In this paradigm, AutoMR is closely related to lifelong LLM-based recommendation systems, which fall into two main categories:
1) Using target items to retrieve similar user sequences, like ReLLa~\cite{rella}, struggles with scalability in generative recommendation scenarios with no candidate set or a very large one. 
2) Using untuned LLMs to summarize users' long-term interests for recommendations, like TRSR~\cite{trsr}. However, LLMs are not specifically designed for recommendation, leading to extracted interests may not be suitable for recommendation systems.
Unlike these methods, AutoMR is designed for generative recommendation scenarios and includes a retrieval module designed and trained specifically for recommendation contexts, demonstrating strong long-term user interest modeling ability in LLM-based generative recommendation.

%% file: main/5_experiments.tex
\section{Experiments}
In this section, we conduct experiments to answer the following research questions:
\begin{itemize}[leftmargin=*]
\item \textbf{RQ1:} Can AutoMR effectively model long-term user interests to improve recommendation quality?
\item \textbf{RQ2:} Are the interactions retrieved by AutoMR more effective than random selections?
\item \textbf{RQ3:} Can AutoMR utilize distant long-term histories for recommendations?
\end{itemize}


\subsection{Experimental Setup}
\subsubsection{Datasets}
We conduct experiments on two representative datasets from Amazon Product Reviews\footnote{\url{https://jmcauley.ucsd.edu/data/amazon/}.}: Amazon Book (Book) and Amazon Movie (Movie). These datasets contain user reviews from Amazon, collected from 1996 to 2018.
Due to the large scale of both Book and Movie, we extract data from 2017 for our experiments, allocating the first 10 months for training, and the remaining two halves for validation and testing, respectively.
We take the reviews as the interaction behaviors and sort the reviews from one user by time to form the behavior sequence.
We set the maximum behavior sequence length to 100, using the most recent 10 behaviors for the short-term sequential features and the rest as the long-term memory bank. To ensure data quality, we filter out items with fewer than 5 interactions and sequences shorter than 20 behaviors.



\subsubsection{Baselines}
We select three famous LLM-based methods as baselines, including 1) \textbf{BIGRec~\cite{bigrec}}, which is an LLM-based generative recommendation model that grounds the output of LLM to actual item space using an L2 distance;
2) \textbf{ReLLa~\cite{rella}}, which models long-term user interest in discriminative recommendations by semantically retrieving the most relevant interacted items to the target item. In generative recommendations, as the target item is unavailable, we adapt ReLLa by using the short-term user behaviors as the query for retrieving; 3) \textbf{TRSR~\cite{trsr}}, which extracts long-term user interest by  
first segmenting user historical behaviors and subsequently utilizing LLMs to summarize long-term user interest from the blocks. Additionally, we choose \textbf{SASRec~\cite{sasrec}}, a representative conventional method, as a baseline. SASRec employs a self-attention mechanism to capture sequential patterns.

\subsubsection{Evaluation metrics}
Following previous works~\cite{sasrec, bigrec}, we leverage the next-item recommendation scheme. 
To evaluate the top-K performance, we employ two widely adopted metrics: Recall, and Normalized Discounted Cumulative
Gain (NDCG). All evaluations are conducted following a full-ranking protocol~\cite{bigrec}.

\subsubsection{Implementation details}
We optimize SASRec using BCE loss and the Adam optimizer, with learning rates searched in [1e-2, 1e-3, 1e-4] and batch sizes in [256, 512, 1,024, 2,048]. For LLM-based methods, we use Llama2-7B as the LLM backbone, optimizing with the AdamW optimizer, exploring learning rates in [1e-3, 1e-4, 1e-5] and setting the batch size to 32. For AutoMR, the retriever is a 2-layer MLP with hidden dimension 256 and we optimize it using the same setting as SASRec. An early stopping strategy is applied based on Recall@1 on the validation set, with a patience of 10. 
\begin{table}[]
\caption{The comparison between AutoMR and baselines.}
\label{table2} 
\begin{tabular}{lcccc}
\toprule
Dataset                & Model   & Recall@1 & Recall@5 & NDCG@5 \\
\hline
\multirow{5}{*}{Book}  & SASRec  & 0.0047   & 0.0214   & 0.0128 \\
                       & ReLLa   & 0.0045   & 0.0065   & 0.0055 \\
                       & TRSR    & \underline{0.0283}   & 0.0351   & 0.0319 \\
                       & BIGRec  & 0.0281   & \underline{0.0360}   & \underline{0.0323} \\ 
\rowcolor[HTML]{E9E9E9}                       & AutoMR & \textbf{0.0291}   & \textbf{0.0379}   & \textbf{0.0338} \\
\hline
\multirow{5}{*}{Movie} & SASRec  & 0.0329   & 0.0554   & 0.0437 \\
                       & ReLLa   & 0.0029   & 0.0050   & 0.0030  \\
                       & TRSR    & \underline{0.0591}   & \underline{0.0637}   & \underline{0.0613} \\
                       & BIGRec  & 0.0575   & 0.0616   & 0.0597 \\ 
\rowcolor[HTML]{E9E9E9}                       & AutoMR & \textbf{0.0601}   & \textbf{0.0638}   & \textbf{0.0621} \\
\bottomrule
\end{tabular}
\end{table}

\subsection{Overall Results (RQ1)}

Table~\ref{table2} illustrates the overall performance comparison between AutoMR and baselines, where we can find:

\begin{itemize}[leftmargin=*, itemsep=2pt, topsep=2pt, parsep=2pt]
    \item AutoMR achieves the best performance. Comparing AutoMR to BIGRec, AutoMR performs better, which verifies 1) the importance of modeling long-term history for generative LLM-based recommendations and 2) the effectiveness of our proposed memory-retrieval framework in modeling long-term interest.
    \item Comparing AutoMR and ReLLa, AutoMR demonstrates superior performance. This is because ReLLa's semantic retrieval can only recall information similar to the current interest, limiting its ability to provide useful signals for prediction. In contrast, AutoMR uses a learning-based approach to train the retriever, enabling it to extract beneficial information from history. 
    \item TRSR outperforms BIGRec on Movie but performs similarly on Book.  We attribute this to the LLM's weaker language understanding of books compared to movies, limiting its ability to extract useful long-term interests from Book data. In contrast, AutoMR overcomes this limitation by leveraging a learning-based paradigm to retrieve effective information from memory.


    
\end{itemize}

\subsection{Analysis}
\subsubsection{Comparison with Random Retrieval (RQ2).} 
\begin{figure}[tbp]
  \centering            
  \subfloat{
        \label{fig:book_wrand}\includegraphics[width=0.225\textwidth]{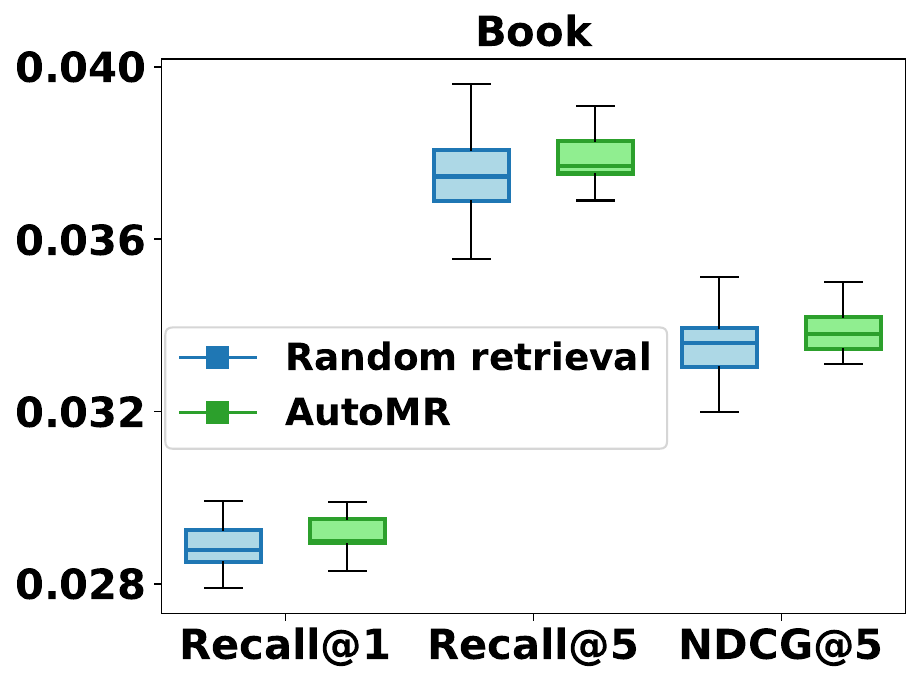}}
   \subfloat{
        \label{fig:movie_wrand}\includegraphics[width=0.225\textwidth]{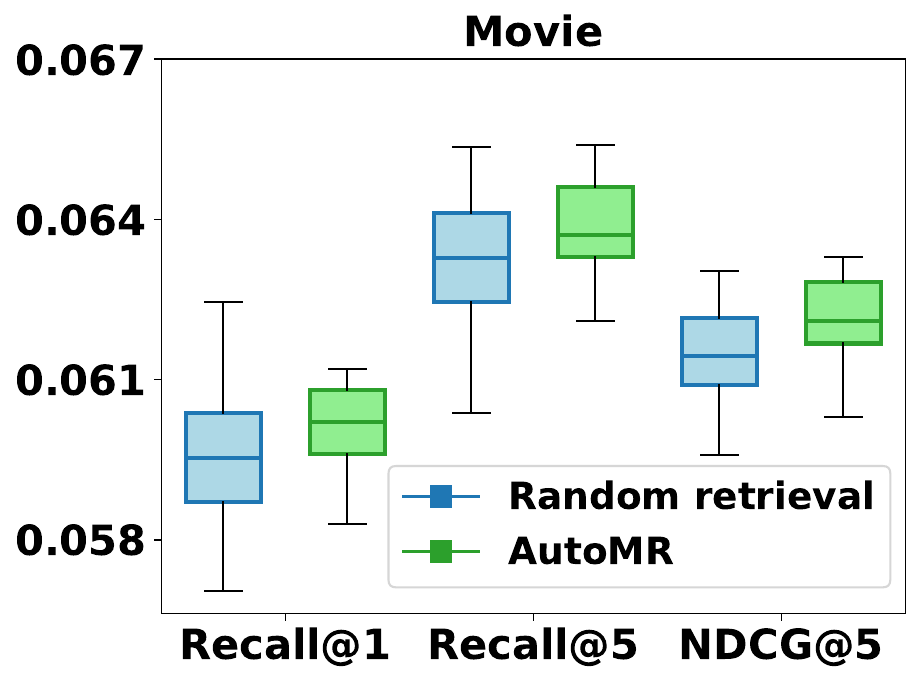}} 
  \caption{Performance comparison of AutoMR and random retrieval. The bottom, middle, and top lines of the box represent the 25th percentile, median (50th percentile), and 75th percentile, respectively.}
  \label{fig:wrand}
\end{figure}

We compare AutoMR with random retrieval and present the results in Figure~\ref{fig:wrand}, where we find AutoMR consistently outperforms random retrieval across both datasets. 
Notably, both the median and 25th percentile values of AutoMR exceed those of random retrieval across all metrics.
This improvement suggests that AutoMR effectively retrieves beneficial long-term history, enhancing future predictions.


\subsubsection{Case Study (RQ3).}
\begin{figure}[tbp]
  \centering            
  \subfloat{
        \label{fig:case_study_book}\includegraphics[width=0.225\textwidth]{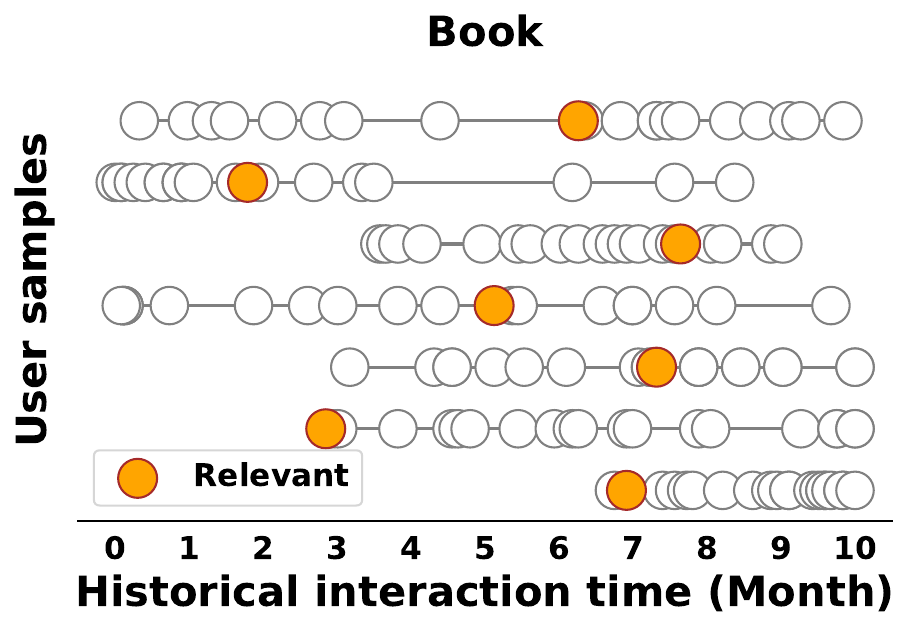}}
  \subfloat{
        \label{fig:case_study_movie}\includegraphics[width=0.225\textwidth]{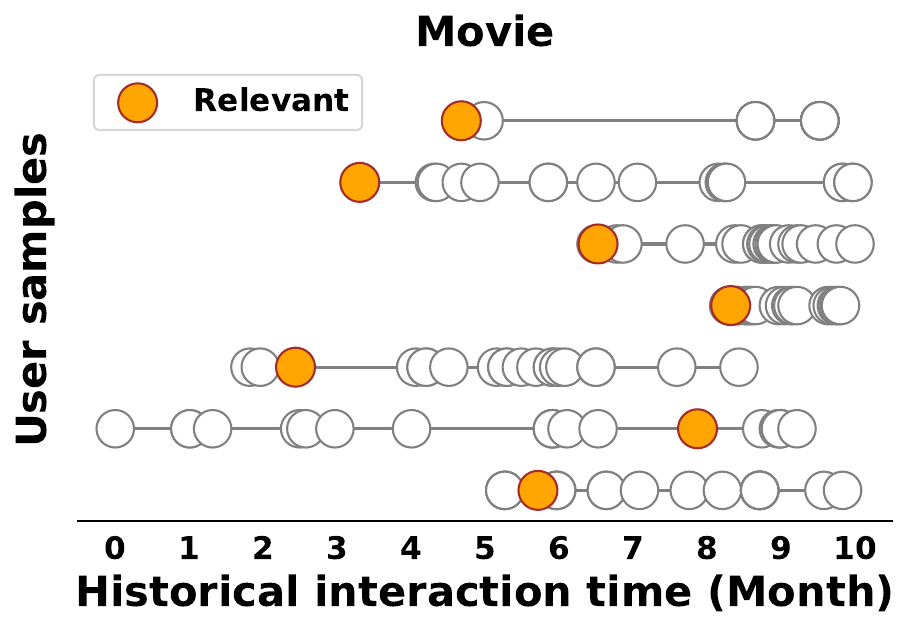}}
  \caption{
Visualization of the historical interaction times for some samples where incorporating long-term history improves Recall@1.
The orange circles represent the long-term history retrieved by AutoMR.
  }
  \label{fig:case_study}
\end{figure}

We analyze the timestamps of long-term histories retrieved by AutoMR for test samples whose Recall@1 improved after incorporating these histories in Figure~\ref{fig:case_study}. 
The figure shows that AutoMR can retrieve users' long-term behaviors even from distant past interactions, highlighting its capability to effectively capture long-term user interests.

%% file: main/6_conclusion.tex
\section{Conclusion}
In this paper, we explored the critical challenge of incorporating users' long-term interests into LLM-based generative recommendation systems. We argue that manually designing an effective retrieval mechanism to identify users' relevant long-term interests is highly challenging.
To address this challenge, we proposed AutoMR, 
that can enhance the well-tuned LLM-based generative recommendation model by training an independent retrieval model to select the most relevant long-term interests.
Our approach leverages a reduction in perplexity (PPL) as a metric to evaluate the relevance of long-term interests, enabling AutoMR to integrate long- and short-term interests effectively. 
Extensive experiments on two widely-used recommendation datasets demonstrate AutoMR's superiority in combining long-term user interest to enhance LLM-based generative recommendation performance against existing baselines.